\documentstyle[12pt]{article}
\title{$\rho\to 4\pi$ DECAY}

\author{S.I. Eidelman, Z.K. Silagadze\\ [1.5mm]
\small\em Budker Institute of Nuclear Physics, 630090, Novosibirsk-90,
Russia\\[1.5mm]
\rm E.A. Kuraev\\
\small \em Joint Institute for Nuclear Research,
141980 Dubna, Moscow region, Russia}
\date{}

\begin{document}
\maketitle
\begin{abstract}

The decay modes $\rho^0 \to 2\pi^+ 2\pi^-$ and $\rho^0 \to2 \pi^0 \pi^+ \pi^-$
are considered in the framework of the low energy effective chiral Lagrangian.
The obtained values of the decay widths $\Gamma(\rho^0\to 2 \pi^+ 2 \pi^-)=$
$(16 \pm 1){\rm keV}$ and $\Gamma(\rho^0\to 2 \pi^0 \pi^+ \pi^-)=
(6.0 \pm 0.2)$keV
do not contradict the existing upper limits and seem to be big enough
for the corresponding processes to be observed in future high luminosity
experiments.
\end{abstract}

      The new generation  experiments at the Novosibirsk
$e^+ e^-$ collider VEPP-2M with two modern detectors \cite{1,2},
as well as the planned experiments at the Frascati $\phi$-factory
DA$\Phi$NE \cite{3} allow  rare decays of the light
vector mesons to be studied. This paper presents results
of the calculation for the
decay widths  $\Gamma(\rho^0 \to 2 \pi^+ 2 \pi^-)$
and
$\Gamma(\rho^0 \to 2 \pi^0 \pi^+ \pi^-)$.
The experimental search  for these decays gave only upper limits
$$\Gamma (\rho^0 \to 2 \pi^+ 2 \pi^-) < 30 \mbox { keV \cite{4}}$$
and
$$\Gamma(\rho^0\to 2 \pi^0 \pi^+ \pi^-) < 6 \mbox { keV \cite{5}}.$$
The significant increase of the total number of $\rho$-mesons expected
in the new  experiments mentioned above motivated this calculation.

Earlier theoretical studies of these decays assumed quasitwoparticle
intermediate states. Renard calculated the width of the decay
$\rho^0\to 2 \pi^0 \pi^+ \pi^-$ via the $\omega \pi$ intermediate state and
obtained the width of 0.9 keV \cite{6} well below the existing
limit, while in \cite{7} the value
$\Gamma(\rho^0\to 2 \pi^+ 2 \pi^-) = 172$ keV considerably higher
than the existing limit  was obtained assuming the
$A_1 \pi$ and $A_2 \pi$ intermediate states. The recent work
\cite{8} pointed to the necessity of taking into account other
intermediate mechanisms within the chiral model and presented the value
of the width of $\Gamma(\rho^0\to 2 \pi^+ 2 \pi^-)$ .
Particularly, they showed that in one of the Yang-Mills type models
(the "Massive Yang-Mills approach" \cite{9}) the value  of 60 keV
is obtained in obvious contradiction to the experimental limit.
They also presented the results for two other models (the Hidden symmetry
scheme and the naive Vector Dominance Model) which are 7.5 and 25 keV
respectively.

 However, it is well known \cite{10,11}
that the simple version of the chiral Yang-Mills Lagrangian as
in \cite{9} should be corrected by special terms so that vector mesons
can be naturally introduced into the Lagrangian without violating
low-energy theorems of the current algebra.
In the present paper such a corrected lagrangian is used \cite{12,13,14}
to calculate the $\rho$-meson decay widths in both channels.

The process $\rho \to 4 \pi$ is in general described by six
classes of Feynman diagrams shown in Fig.~1.
The square of the  corresponding matrix element averaged over  spin states
is given by the following formula:
\begin{eqnarray}
 \overline{\vert M \vert ^2}={1\over 3}g^2_{\rho\pi\pi}
\vec{J}^*(\rho \to 4\pi) \cdot \vec{J}(\rho \to 4\pi) \hspace*{5mm},
\label{eq1} \end{eqnarray}
where $g_{\rho\pi\pi}J_\mu(\rho \to 4 \pi)$ is the conserved current
which describes the $\rho \to 4\pi$ transition.

\noindent
.\hspace{-2mm}\special{em:graph rho.pcx}

\vspace{54mm}
\noindent
Fig. 1. Feynman diagrams describing the $\rho\to 4\pi$ decay.

\vspace{4mm}
 For the reaction
 $\rho^0(q) \to \pi^+(q_1) + \pi^+(q_2) + \pi^-(q_3)+\pi^-(q_4)$
  only diagrams a-d from Fig.~1 contribute.
 The corresponding current has a form:
\begin{eqnarray}
&& J_\mu^{\rho^0\to 2\pi^+ 2\pi^-} = \left({1\over 3}-\alpha_k\right)
{1\over f_\pi^2}
\left[ 6 (q_1 + q_2 - q_3 - q_4)_\mu  \right. \nonumber \\
&& + (6 q_3.q_4 + 2 m^2)
\left({(q - 2 q_1)_\mu\over (q - q_1)^2 - m^2}
+ {(q - 2 q_2)_\mu\over (q - q_2)^2 - m^2}\right) \nonumber\\
&&-\left. ( 6 q_1.q_2 + 2 m^2)\left({(q - 2q_3)_\mu
\over (q - q_3)^2 - m^2} + {(q - 2 q_4)_\mu\over (q - q_4)^2 - m^2}
\right)\right]  \nonumber\\
&& + 2 (1 + P_{12}) (1 + P_{34}) {g_{\rho\pi\pi}^2((q_2 + q_4)^2 - m_{\rho}^2
- i m_\rho \Gamma_\rho) \over ((q_2 + q_4)^2 - m_\rho^2)^2
+ m_\rho^2\Gamma_\rho^2} \nonumber\\
&&\times \left[ (q_4 - q_2)_\mu+ {q_1.(q_2 - q_4)\over
(q - q_3)^2 - m^2} (q - 2 q_3)_\mu + {q_3.(q_2 - q_4)\over (q - q_1)^2
- m^2} (q - 2 q_1)_\mu\right],
\label{eq2}
\end{eqnarray}
\noindent
where $m^2=m_{\pi^\pm}^2=q_i^2$,$\, \alpha_k={f^2_\pi g^2_{\rho\pi\pi}\over
m^2_\rho}
\approx 0.55$, and $P_{12}$ and $P_{34}$ operators stand for
the interchange of the momenta of the corresponding identical mesons.

For the process
$\rho^0(q) \to \pi^+(q_+) + \pi^-(q_-) + \pi^0(q_1) + \pi^0(q_2)$
all six classes of diagrams contribute.
One of them (f) contains the $\omega \pi$ intermediate state and is due to
the anomalous part of the chiral Lagrangian.

 The corresponding current $J_\mu$ can be presented as a sum
of three terms, each of them representing a gauge invariant subset
of diagrams ($J_\mu^{(i)}q^\mu=0$):
\begin{eqnarray}
J_\mu^{\rho^0\to 2\pi^0 \pi^+ \pi^-} = J_\mu^{(1)} + J_\mu^{(2)} + J_\mu^{(3)}.
\label{eq3}\end{eqnarray}
Diagrams of type a,b of Fig.~1 give after some algebraic transformations:
\begin{eqnarray}
J_\mu^{(1)}& =& \left( \frac{1}{3} -\alpha_k\right) \frac{1}{f_\pi^2}
( 6 q_1\cdot q_2 + 2 m_{\pi^0}^2) \\
&\times& \left[ \frac{(q - 2 q_-)_\mu}{(q - q_-)^2 -
m_{\pi^\pm}^2 }\ - \frac{(q - 2 q_+)_\mu}{(q - q_+)^2 -
m_{\pi^\pm}^2}\right] .\nonumber
\label{eq4} \end{eqnarray}
The second piece arises from diagrams of type c,d,e of Fig.~1 and has
the form
\begin{eqnarray}
J_\mu^{(2)}&&\hspace{-3mm}=- g_{\rho\pi\pi}^2\, (1+P_{12})\, \left\{ -
\frac{1}{r_+r_-} \bigl[ 2 (q_+ - q_1)_\mu \, q\cdot (q_- - q_2) \right. \bigr.
\nonumber \\ &&\hspace{-3mm}- \bigl. 2 (q_- - q_2)_\mu \,
q\cdot (q_+ - q_1) + (q_2 + q_- - q_1 - q_+)_\mu \, (q_+ - q_1) \cdot (q_- -
q_2)\bigr] \nonumber \\
&&\hspace{-3mm}+ \frac{1}{r_+}\, \left[ (q_+ - q_1)_\mu - 2 q_2\cdot (q_+ -
q_1)
\frac{(q - 2 q_-)_\mu}{(q - q_-)^2 - m_{\pi^\pm}^2}\right] \nonumber \\
&&\hspace{-3mm}- \left. \frac{1}{r_-}\, \left[ (q_- - q_2)_\mu - 2 q_1\cdot
 (q_- - q_2)
\frac{(q - 2 q_+)_\mu}{(q - q_+)^2 - m_{\pi^\pm}^2 } \right]\right\} ,
\label{eq5} \end{eqnarray}

\begin{eqnarray*}
&&r_+ = (q_+ +q_1)^2 - m_\rho^2 + i m_\rho \Gamma_\rho ;\\
&&r_- = (q_- + q_2)^2 - m_\rho^2 + i m_\rho \Gamma_\rho .
\end{eqnarray*}
Finally, the third part of the current is determined by two diagrams
of type f of Fig.~1 with the $\omega$-meson intermediate state
(note that the $\omega \to 3\pi$ effective vertex contains
contributions from the contact term as well as from the $\rho \pi$
intermediate states):
\begin{eqnarray}
J_\mu^{(3)} = {3 g_{\rho\pi\pi} \over 8\pi^2 f_\pi}\, (1+P_{12})\,
P_\mu \, {F_1\over r_1},
\label{eq6} \end{eqnarray}
where
\begin{eqnarray}
P_\mu &=& q_1\cdot q_2 (q_{+\mu} q\cdot q_- - q_{-\mu} q\cdot q_+) +
q_-\cdot q_2 (q_{1\mu}
q\cdot q_+ - q_{+\mu} q\cdot q_1) \\
&+& q_+\cdot q_2 (q_{-\mu}
q\cdot q_1 - q_{1\mu} q\cdot q_-),\nonumber
\label{eq7}\end{eqnarray}
and
\begin{eqnarray}
&&r_1\>\> = (q-q_2)^2 - m_\omega^2 + i m_\omega \Gamma_\omega \ , \\[2mm]
&&F_1\>\> = {3g_{\rho\pi\pi} \over 4 \pi^2 f_\pi^3}
\left[1-3\alpha_k -\alpha_k\left({m_\rho^2\over r_{+-}} +
{m_\rho^2\over r_{+1}} + {m_\rho^2\over r_{-1}} \right)\right],\nonumber\\[2mm]
&& r_{+-} = (q_+ + q_-)^2 - m_\rho^2 + i m_\rho \Gamma_\rho, \nonumber\\
&& r_{+1}\> = (q_+ + q_1)^2 - m_\rho^2 + i m_\rho \Gamma_\rho, \nonumber\\
&& r_{-1}\> = (q_1 + q_-)^2 - m_\rho^2 + i m_\rho \Gamma_\rho.\nonumber
\label{eq8}\end{eqnarray}
\noindent
   The expressions for the current      above use the values
   of the constants from \cite{13}.

   The width of the decay is given by the following expression
\begin{eqnarray}
\Gamma=\frac{N}{2m_\rho(2\pi)^8}\,R
\nonumber \end{eqnarray}
\noindent
 where a factor   $N$ takes into account the identity of the final
pions and equals 1/4 for $2\pi^+2\pi^-$ and 1/2 for $2\pi^0\pi^+\pi^-$.
$R$ is the above matrix element squared  (formula (1)) integrated over the
phase space of final particles. This integration  was performed
by two independent methods. In one of them the quantity $R$
was represented as the following five-dimensional integral \cite{15}
and was calculated numerically:
\begin{eqnarray}
&&.\hspace{-5mm}R={\pi^2\over 8 m^2_\rho}\, \int_{s_1^-}^{s_1^+} d s_1
\int_{s_2^-}^{s_2^+} d s_2  \int_{u_1^-}^{u_1^+}
{d u_1 \over \sqrt{\lambda(m^2_\rho,s_2,s_2^\prime)}}\nonumber\\
&&\times \int_{u_2^-}^{u_2^+} d u_2
\int_{-1}^{1} {d\zeta \over \sqrt{1-\zeta^2}} \overline{|M|^2}.
\nonumber \label{eq9}\end{eqnarray}
Here we introduced Kumar's invariant variables
$$s_1=(q-q_1)^2,\  s_2=(q-q_1-q_2)^2,\  u_1=(q-q_2)^2, $$
$$ u_2=(q-q_3)^2,\ t_2=(q-q_2-q_3)^2,$$
$s_2^\prime = s_2+s+m_1^2+m_2^2-u_1-s_1$ and ${\rm arccos}\zeta$ is an
angle between $(\vec q_2, \vec q_1 + \vec q_2)$ and $(\vec q_3, \vec q_1 +
\vec q_2)$ planes. $\lambda(x,y,z)=(x+y-z)^2-4xy$ is a conventional triangle
function. The relation between $t_2$ and $\zeta$ as well as
the expressions for the integration limits can be found in \cite {15}.

Another method used the Monte-Carlo procedure of the random star generation
suggested by Kopylov \cite{16}. Both methods gave consistent
results.

   The  values of the widths obtained were
$$\Gamma(\rho^0 \to 2\pi^+2\pi^-)=(16 \pm 1)\ {\rm keV} $$
$$\Gamma(\rho^0 \to \pi^+\pi^-2\pi^0) = (6.0\pm 0.2)\ {\rm keV} $$
\noindent
and are close to  the existing upper limits but
do not contradict them.

   From these results one can estimate the peak values of the
cross section of four pion production in $e^+e^-$ annihilation
 in the vicinity of the $\rho$-meson resonance.
   The obtained values are  0.12nb   and   0.04nb  respectively and give
a real chance of observing these processes in the forthcoming
experiments.

    We have also calculated the width of a similar decay
  $\phi \to \eta\pi^+\pi^-\pi^0$. Unfortunately, the  value of
   the width obtained
  corresponds to a very small branching ratio of about $10^{-11}$ and
  can hardly be observed in the near future.

\end{document}